\documentclass[aps,amsmath,twocolumn,amssymb,floatfixng,showpacs,
superscriptaddress,footinbib, longbibliography]{revtex4-1}

\usepackage{graphicx}% Include figure files
\usepackage{dcolumn}% Align table columns on decimal point
\usepackage{multirow}
\usepackage{booktabs}
\usepackage{bm,color}
\usepackage{braket}
\usepackage{amsmath,amssymb}
\usepackage[colorlinks,linkcolor=blue,hyperindex,CJKbookmarks]{hyperref}
\usepackage{epstopdf}
\usepackage{gensymb}

\definecolor{red1}{rgb}{0.6,0,0}

\newcommand{\bgea}{\begin{equation}}
\newcommand{\enea}{\end{equation}}
\newcommand{\bea}{\begin{eqnarray}}
\newcommand{\eea}{\end{eqnarray}}

\begin{document}

\title{Correlation enhanced altermagnetism mediated by spin-lattice coupling in CrSb}

\author{Charmi Bhalani}
\affiliation{Department of Physics, École Centrale School of Engineering, Mahindra University, Hyderabad-500043, India}
\author{Banasree Sadhukhan}
\email{banasree.sadhukhan@mahindrauniversity.edu.in}
\affiliation{Department of Physics, École Centrale School of Engineering, Mahindra University, Hyderabad-500043, India}

\begin{abstract}

Altermagnets (AMs) exhibit momentum-dependent spin splitting without net magnetization, making them promising platforms for spintronic applications. While the symmetry and multipolar origins of nonrelativistic altermagnetic spin splitting (NRASS) are well established, the role of electronic correlations and spin-lattice coupling (SLC) in controlling NRASS remain unexplored. In particular, SLC provides a direct pathway for tuning altermagnetism through lattice degrees of freedom. Here we first investigate the role of electronic correlations on altermagnetic strength in NiAs-type material CrSb using density functional theory (DFT), DFT+$U$, and dynamical mean-field theory (DMFT). We find that increasing electronic correlations substantially enhance NRASS and drive the stabilization of an incommensurate spin-spiral (SS) state, while dynamical correlations further amplify the spin splitting through quasiparticle renormalization of the Cr-$3d$ states. We find that the NRASS remains 12.5\% larger in DFT+DMFT than in DFT. By evaluating SLC in next step, we establish its direct connection to the evolution of NRASS and identify SLC as a microscopic descriptor between electronic correlation and enhanced altermagnetism in CrSb. Investigating it in another NiAs-type compound MnTe, we show that the SLC-NRASS correlation is generic across AMs and establish SLC as a microscopic descriptor of altermagnetic strength. Our results provide a unified framework for tuning altermagnetism through the interplay between the electronic correlations and lattice degrees of freedom.

\end{abstract}

\maketitle

%###################################

\section{Introduction}

\par The discovery of altermagnetism has established a new paradigm in magnetic materials by revealing a class of compensated magnets that exhibit momentum-dependent spin splitting despite possessing zero net magnetization \cite{PhysRevX.12.031042, PhysRevB.110.054406, PhysRevLett.132.036702, Krempasky2024, PhysRevX.12.040501, PhysRevLett.132.176702, PhysRevX.12.011028, 10.1039/d6ra02985j}. In contrast to conventional antiferromagnets, altermagnets (AMs) combine the absence of macroscopic magnetization with spin-polarized electronic states, thereby merging desirable features of antiferromagnets and ferromagnets \cite{PhysRevMaterials.5.014409, PhysRevX.12.031042, PhysRevX.12.011028}. The resulting spin splitting originates from the crystal and magnetic symmetries of the system rather than from spin-orbit coupling (SOC) offering a robust mechanism for generating spin-polarized electronic responses \cite{PhysRevX.12.031042, doi:10.1126/sciadv.aaz8809, PhysRevLett.132.056701}. These unique properties have stimulated intense interest in altermagnetic materials as promising platforms for next-generation spintronic technologies \cite{Fu2025, j3qj-77yj, Reichlova2024, bai2024altermagnetism, Mukherjee_2026, Pournaghavi2025-ul}. In addition to the spin degree of freedom, their electronic and magnetic properties can be tuned through orbital and lattice degrees of freedom which provides additional pathways for controlling altermagnetic behavior \cite{l8fc-dp36, rkr4-p5n4}.

\par Recent theoretical and experimental studies have identified several AMs, including MnTe \cite{PhysRevB.107.L100418, PhysRevLett.132.036702, PhysRevMaterials.8.L041402, PhysRevB.109.115102, dzk6-f68q}, MnF$_2$ \cite{PhysRevLett.134.226702}, and CrSb \cite{Reimers2024, Yang2025} where opposite spin sublattices are correlated by the rotational symmetry. Among these, CrSb has emerged as a particularly appealing system owing to its simple crystal structure and sizable altermagnetic spin splitting near the Fermi level \cite{PhysRevLett.133.206401, Yu2025-bt, PhysRevMaterials.8.084412}. Previous investigations have primarily focused on its collinear antiferromagnetic ground state, where the characteristic momentum-dependent spin polarization has been shown to give rise to unconventional transport and spin-dependent phenomena. However, the stability of altermagnetic order beyond the collinear regime  remain largely unexplored  \cite{r34k-xjpx, kq6x-7jfc, Cheong2024, Hu2025}.

\par Electronic correlations play a central role in determining the magnetic and electronic properties of transition-metal compounds. In Cr-based systems, local Coulomb interactions can significantly modify exchange interactions, spin and orbital magnetic moments, and low-energy electronic states \cite{PhysRevB.102.115162}. Such correlation effects may alter not only the magnitude of the  non-relativistic altermagnetic spin splitting (NRASS) but also the nature of the magnetic ground state itself \cite{pgp6-zlh8,Len2025, PhysRevB.65.184435, PhysRevB.110.174412}. In particular, enhanced electronic correlations can promote competing magnetic interactions, potentially driving a transition from a collinear antiferromagnetic state to more complex noncollinear spin textures \cite{PhysRevB.111.205429}. Understanding how magnetic reconstruction governs altermagnetic behavior is therefore of fundamental importance \cite{kq6x-7jfc, Cheong2024, Hu2025}.

\par Beyond electronic correlations, the coupling between spin and lattice degrees of freedom provides an additional mechanism for tuning magnetic and electronic properties \cite{PhysRevB.105.104418, SciPostPhys.18.2.064, rb2d-qzqs, PhysRevMaterials.9.024409}. Spin-lattice couplings (SLCs) describes the dependence of magnetic interactions on atomic displacements due to lattice vibration and local structural distortions. This varies the bond lengths, bond angles, and crystal-field environments which modifies the underlying exchange interactions \cite{PhysRevB.105.104418}. In particular, both the isotropic Heisenberg exchange and the relativistic Dzyaloshinski-Moriya interactions (DMIs) can be highly sensitive to lattice distortions \cite{PhysRevB.105.104418, SciPostPhys.18.2.064}, leading to substantial changes in the stability of competing magnetic states. Recent first-principles studies have further shown that the microscopic origin of strong SLC can be traced to the competition among orbital-dependent exchange channels, whose relative strengths are controlled by the local bonding geometry \cite{PhysRevB.105.104418, SciPostPhys.18.2.064, rb2d-qzqs, PhysRevMaterials.9.024409}.

\par Previous studies have established a multipolar framework that correlates the magnitude of NRASS with specific magnetic multipoles. It provides a microscopic foundation for understanding and tuning the altermagnetic response \cite{PhysRevResearch.6.043157,1r6k-s46h,PhysRevX.14.011019,PhysRevB.88.094429}. More recently, this framework has been combined with phonon-assisted approaches to demonstrate how symmetry-resolved lattice distortions and magnetic dipolar order govern the emergence and therefore modulation of NRASS \cite{1r6k-s46h,https://doi.org/10.1002/adts.202501879,c3kh-5s5t}. Despite these recent advances, the interplay between electronic correlations and lattice degrees of freedom in determining NRASS remains largely unexplored. Motivated by this recent challenge, the key question we address here is that can correlation-induced modifications of magnetic exchange and SLC provide a microscopic mechanism for controlling the altermagnetic spin splitting? More specifically, we ask whether the correlation dependence of SLC can mediate the evolution of NRASS by coupling electronic correlations to symmetry-selected lattice distortions and magnetic interactions. Establishing such a mechanism would bridge the two separate descriptions of correlation effects. One is lattice dynamics and another one is altermagnetism. Therefore it will provide a microscopic framework for understanding and tuning correlation-driven altermagnetic responses mediated by SLC.

\par  We investigate the evolution of altermagnetic strength in CrSb using a combination of  density functional theory (DFT), DFT+$U$, and dynamical mean-field theory (DMFT). We demonstrate that increasing electronic correlations substantially enhance the NRASS and drive a correlation-induced transition from the collinear antiferromagnetic state to a spin-spiral state in CrSb. DMFT calculations further reveal that dynamical electronic correlations reinforce this enhancement through quasiparticle renormalization of the Cr-$3d$ states. By explicitly evaluating the SLCs, we uncover its direct quantitative correlation with the evolution of NRASS and establish SLC as the microscopic mechanism linking electronic correlations to the enhanced altermagnetic response via lattice degrees of freedom. Furthermore, a comparative analysis of CrSb and the prototypical NiAs-type another AM MnTe demonstrates that this correlation is generic across distinct altermagnetic systems and identifies SLC as a universal microscopic descriptor of altermagnetic strength.

%###################################

\section{computational details}

\par DFT calculations are performed using the full potential muffin-tin orbital based Relativistic Spin Polarized toolkit (RSPt) \cite{rsptweb} code within the local spin density approximation (LSDA), extended with DFT+U and DFT+DMFT to capture electronic correlation effects in CrSb. A k-point mesh 12 $\times$ 12 $\times$ 12 is used for without SOC. While a denser k-point mesh 16 $\times$ 16 $\times$ 16 is used for with SOC calculation. Considering the magnetic exchange tensor $\mathrm{J}^{\alpha\beta}_{ij}$ as a $3\times3$ matrix, the isotropic (Heisenberg) exchange interaction $\bar{J}_{ij}$ and the magnitude of the antisymmetric DMI, $D_{ij}$, are defined as
\begin{eqnarray}
 {\mathrm{\bar J}}_{ij} &=& \frac{\mathrm{J}_{ij}^{xx}+\mathrm{J}_{ij}^{yy}+\mathrm{J}_{ij}^{zz}}{3}, \nonumber\\
D_{ij} &=& |{\bm {\mathrm{D}}}_{ij}| =
\sqrt{(D_{ij}^{x})^2+(D_{ij}^{y})^2+(D_{ij}^{z})^2},
\label{def-int}
\end{eqnarray}
where  $|{\bm {\mathrm{D}}}_{ij}|$ denotes the magnitude of the DMI vector whose $z$-component is given by
\bea
D^z_{ij} = (J^{xy}_{ij} - J^{yx}_{ij})/2. %\nonumber
\eea
Here, $\alpha,\beta=x,y,z$ represent the Cartesian coordinates, while $i$ and $j$ denote the atomic indices. The exchange interaction tensors ($\mathrm{J}^{\alpha\beta}_{ij}$) are calculated using the magnetic force theorem implemented within RSPt \cite{rsptweb} for both DFT+U and DFT+DMFT. Green's function technique is used within linear response theory with a small angular deviations of the magnetic moments from their equilibrium directions. The tensor components are obtained from the second-order response to these deviations \cite{antropov1997exchange,PhysRevB.68.104436,PhysRevB.79.045209,secchi2015magnetic}.

\par We adopt the convention that positive  ${\mathrm{\bar J}}_{ij}$ corresponds to ferromagnetic (FM) exchange, whereas negative  ${\mathrm{\bar J}}_{ij}$ corresponds to antiferromagnetic (AFM) exchange. Using the calculated  ${\mathrm{\bar J}}_{ij}$ and  ${\bm {\mathrm{D}}}_{ij}$, we construct the corresponding spin Hamiltonian given by :
\begin{eqnarray}
\begin{aligned}
    {\mathrm{H}_s}= -\sum_{i,j} {\mathrm {\bar J}}_{ij} {\bm {\mathrm{S}}}_i \cdot {\bm {\mathrm{S}}}_j - \sum_{i,j} {\bm {\mathrm{D}}}_{ij} \cdot ({\bm {\mathrm{S}}}_i \times {\bm {\mathrm{S}}}_j)\nonumber
 \end{aligned}   
\label{eq1}
\end{eqnarray}
and investigate the magnetic ground state using the spin Hamiltonian above through a combination of both spin-dynamics (SD) and Monte Carlo (MC) simulations \cite{UppASD_book,uppasd}.
We took 25$\times$25$\times$50 supercell in both MC and SD simulation with a Gilbert damping factor of 0.1 in Landau-Lifshitz Gitbert equation \cite{PhysRevB.94.144419}. We employ heat-bath MC annealing before SD simulations to ensure the correct magnetic ground state of the spin Hamiltonian \cite{PhysRevB.34.6341} for both DFT+U and DFT+DMFT. In the heat-bath MC annealing, the simulations are initialized from a random spin configuration. The system is then annealed in 7 temperature steps from $T=500$ K to $T=0.0001$ K. At each temperature, $5\times10^{6}$ MC steps are used for thermalization. The measurement phase is performed at $T=0.0001$ K using an additional $2\times10^{6}$ MC steps to identify the lowest-energy magnetic configuration.

\par Phonon dispersion calculations are performed within DFT using the Quantum ESPRESSO-7.3 package with PBE pseudopotentials \cite{QuantumESPRESSO, giannozzi2017advanced, giannozzi2009quantum}. The self-consistent calculations employ a plane-wave energy cutoff of 65 Ry and a $12\times12\times8$ Monkhorst-Pack $k$-point mesh, with Gaussian smearing of 0.005 Ry. The Kohn-Sham equations are solved using conjugate-gradient diagonalization with a mixing parameter of 0.5, achieving a total-energy convergence threshold of $10^{-9}$ Ry. Phonon bands are calculated within density functional perturbation theory (DFPT) using a $4\times4\times4$ $q$-point mesh covering the full Brillouin zone (BZ), with a self-consistency threshold of $10^{-14}$.

\section{Electronic and magnetic structure of CrSb}

\par
\begin{figure}  
\includegraphics[width=0.5\textwidth,angle=0]{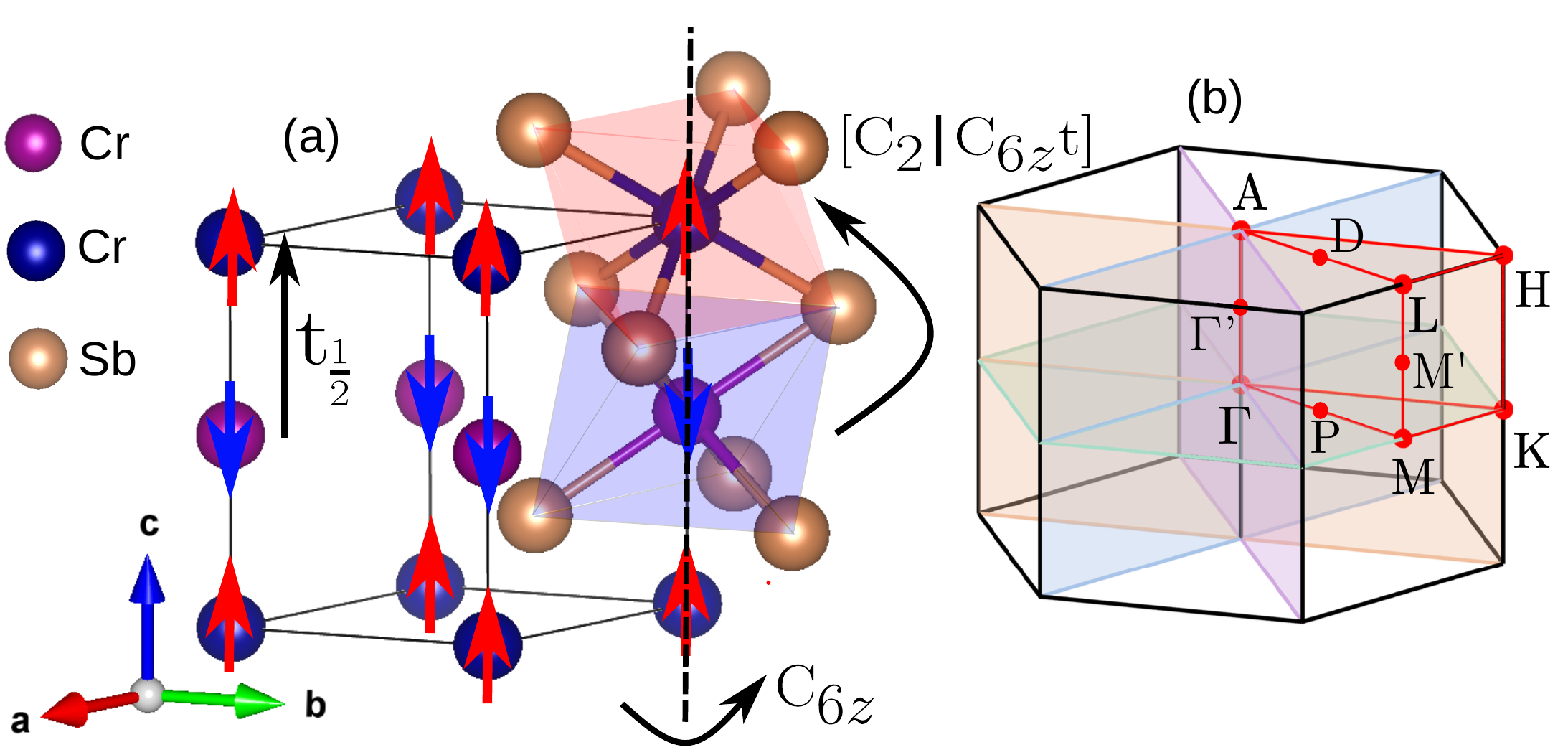} 
\caption{(a) Crystal structure of CrSb with the up- and down-spin sublattices distinguished by red and
blue shading. The sublattices are connected by a non-symmorphic sixfold screw-axis rotation along c-axis leading to the opposite spin sublattices are connected via combination of rotation
and translation operation in real space. (b) Nodal planes of CrSb in three dimensional Brillouin zone showing high symmetry kpoints.}
\label{fig1} 
\end{figure}

 CrSb crystallizes in the hexagonal NiAs-type structure with space group $P6_{3}/mmc$ (No. 194) with lattice parameters a = 4.1243 \AA\ and c = 5.4728 \AA\, where Cr atoms occupy the $2a$ Wyckoff positions and Sb atoms reside at the $2c$ sites \cite{PhysRevLett.133.206401}. The crystal structure consists of alternating Cr and Sb layers stacked along the $c$ axis, forming a three-dimensional network of edge-sharing CrSb$_6$ octahedra. The two sublattices are connected by a nonsymmorphic sixfold screw-axis operation along the c-axis (C$_{6z}$), which relates the opposite spin sublattices through a combined rotational and translational symmetry operation ($[C_{2}\,|\,C_{6z}t_{1/2}]$) in real space as shown in Fig.\ref{fig1}(a). Figure \ref{fig1}(b) represents three dimensional BZ zone along with nodal lines and high symmetry kpoints in NiAs type family. The magnetic properties are predominantly governed by the partially filled Cr-$3d$ orbitals, while the Sb-$5p$ states contribute mainly to the electronic structure near the Fermi level through substantial $p$--$d$ hybridization.

\begin{figure*}  
\includegraphics[width=0.95\textwidth,angle=0]{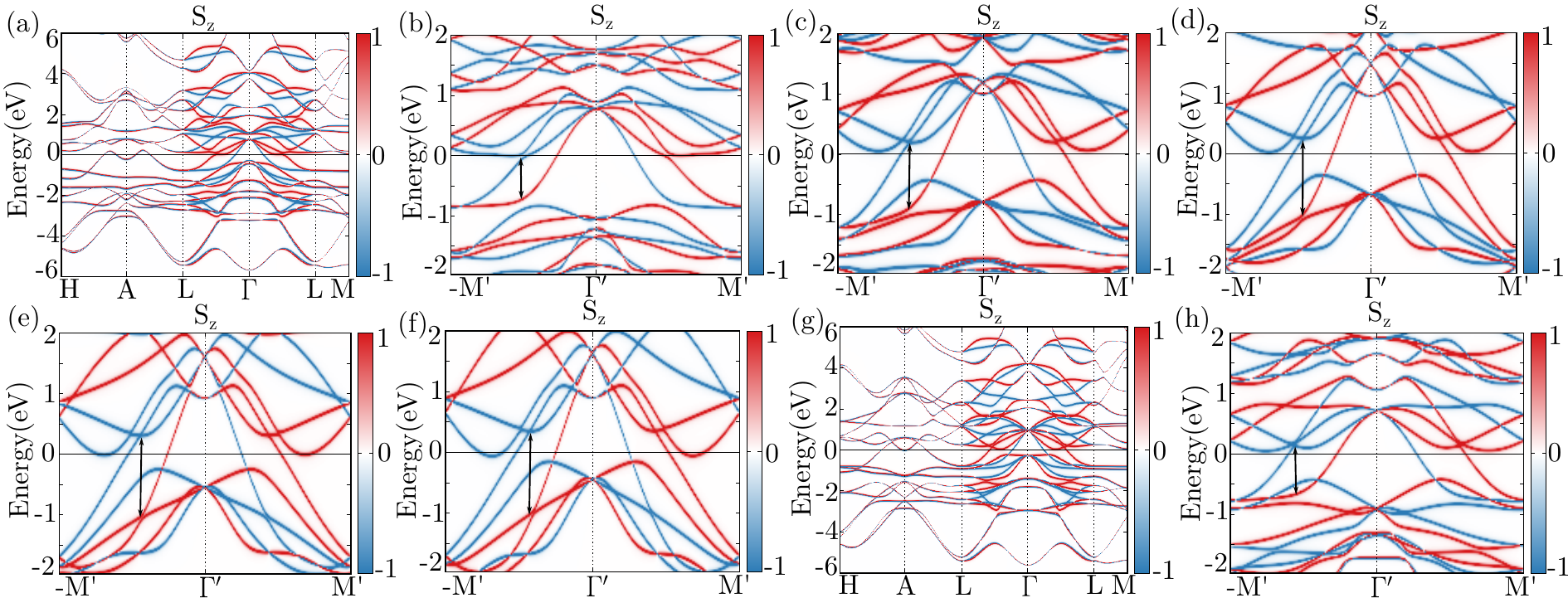} 
\caption{Non-relativistic spin-split band structures for CrSb without SOC along the H-A-L-$\Gamma$-L-M and -$M'$–$\Gamma'$–$M'$ paths from DFT+$U$ with (a)–(b) $U=0$, (c) $U=1$, (d) $U=2$, (e) $U=3$, and (f) $U=4$ eV, and from (g)–(h) DFT+DMFT respectively.}
\label{fig2} 
\end{figure*}

\par To establish the magnetic ground state, we performs spin-polarized density-functional calculations for several collinear magnetic configurations. The AFM state is found to be energetically favorable over the nonmagnetic and FM configurations, consistent with previous experimental and theoretical studies \cite{PhysRevB.102.224426, 10.1021/acs.nanolett.5c00482, k9zm-2k1v, PhysRevLett.133.206401}. In the AFM phase, the Cr moments are aligned parallel within the basal plane and antiparallel between neighboring magnetic sublattices, resulting in a compensated magnetic structure with a vanishing net magnetization. The calculated local magnetic moment on each Cr atom is approximately $m_{\mathrm{Cr}} \sim$ {\textcolor{blue} 2.17 }$\mu_{\mathrm{B}}$  from DFT calculations (see table \ref{table-1}), indicating the localized nature of the Cr-$3d$ electrons.

\par We first examine the non-relativistic electronic band structure, which reveals the characteristic altermagnetic spin splitting along selected high-symmetry directions in BZ. Figure~\ref{fig2} (a) shows the electronic band structure of AFM CrSb, calculated without SOC, along the high-symmetry path H--A--L--$\Gamma$--L--M. A pronounced altermagnetic spin splitting is observed exclusively along the $\Gamma$--L--M direction, whereas the remaining high-symmetry paths retain spin degeneracy.

\par The low-energy electronic structure is predominantly derived from Cr-$3d$ orbitals, while the Sb-$5p$ states contribute over a wider energy window below the Fermi level. Significant hybridization between the Cr-$3d$ and Sb-$5p$ orbitals results in several dispersive bands crossing the Fermi energy, confirming the metallic nature of AFM CrSb. Importantly, although the system possesses zero net magnetization, the electronic bands exhibit a distinct momentum-dependent spin splitting along L-$\Gamma$--M direction, which is a hallmark of altermagnetism. This unconventional spin splitting arises from the interplay between the crystal symmetry and the collinear AFM order, producing opposite spin polarizations at symmetry-related momentum points while maintaining complete compensation of the magnetic moments.

\begin{table}
\centering
\begin{tabular}{|c|c|c|}
\hline
Calculation & Magnetic moment($\mu$) & $\Delta$E  \\
\hline
DFT & 2.17 & 0.72 \\
\hline
DFT+U=1 & 2.86 & 1.05 \\
\hline
DFT+U=2 & 3.18 & 1.21 \\
\hline
DFT+U=3 & 3.40 & 1.31 \\
\hline
DFT+U=4 & 3.55 & 1.32 \\
\hline
DFT+DMFT & 2.28 & 0.81 \\
\hline
\end{tabular}
\caption{Calculated magnetic moment and maximum non-relativistic
altermagnetic spin splitting along M$^{'}$--$\Gamma^{'}$--M$^{'}$ direction in Brillouin zone from DFT+U with $U=1$--4 eV and DMFT for CrSb.}
\label{table-1}
\end{table}

%###################################

\section{Correlation-Driven Enhancement of Altermagnetic Spin Splitting}

\par To examine the influence of electronic correlations, we performed DFT+$U$ and DMFT calculations by incorporating an on-site Coulomb interaction for the Cr-$3d$ orbitals. The evolution of the magnetic moment with increasing electronic correlation strength is summarized in table~\ref{table-1}. The inclusion of the Hubbard $U$ strengthens the exchange splitting of the Cr-derived bands and redistributes the spectral weight in the vicinity of the Fermi level, while preserving the metallic character of the system. As demonstrated in the following sections, the impact of electronic correlations becomes substantially more pronounced, leading to a significant enhancement of NRASS. These results underscore the crucial role of electron correlations in reinforcing the altermagnetic electronic structure of CrSb.

\par The magnitude of the NRASS is primarily governed by the exchange interactions and the orbital-dependent electronic structure in the vicinity of the Fermi level. Given the moderately correlated nature of the Cr-$3d$ electrons in CrSb, it is essential to investigate the influence of electronic correlations on the altermagnetic response. Our DFT calculations yield a maximum NRASS of 0.72 eV along the M$^{'}$--$\Gamma^{'}$--M$^{'}$ direction, as indicated by the arrow in Fig.~\ref{fig2}(b). The M$^{'}$-$\Gamma^{'}$-M$^{'}$ path lies in the $k_z=\pi/2c$ plane, whereas the M-$\Gamma$-M and L-A-L paths are located in the $k_z=0$ and $k_z=\pi/c$ planes, respectively (see Fig.\ref{fig1}(b)). To elucidate the role of electronic correlations, we systematically examined the evolution of the NRASS within the DFT+$U$ framework by varying the on-site Hubbard Coulomb interaction over the range $U=1$-4 eV on Cr-$3d$ states. Furthermore, to capture dynamical many-body effects beyond the static mean-field approximation, we carried out DFT+DMFT calculations and compared the resulting spin-splitting behavior with the DFT and DFT+$U$ results.

\par Figure~\ref{fig2}(c)-(f) present the electronic band structures obtained for different values of $U$ along M$^{'}$-$\Gamma^{'}$-M$^{'}$ path. While the overall band topology remains largely unchanged, a pronounced increase in the momentum-dependent NRASS is observed with increasing Coulomb interaction. The enhancement is particularly significant for the Cr-$3d$-dominated bands located near the Fermi energy, where correlation effects strengthen the effective exchange field experienced by itinerant electrons. To quantify this behavior, we extracted the maximum altermagnetic spin splitting, $\Delta E$, along representative high-symmetry directions. $\Delta E$, which is the magnitude of NRASS, increases monotonically with increasing $U$ as shown in table \ref{table-1}. The energy splitting $\Delta E$ increases to 1.32 eV within DFT+$U$ ($U=4$ eV), representing an 83.33\% enhancement compared to the DFT value. The enhancement originates from the increased localization of the Cr-$3d$ orbitals, which amplifies the exchange interactions while preserving the underlying crystal symmetries responsible for altermagnetic order. Consequently, electronic correlations do not alter the symmetry-enforced nature of the spin splitting but substantially strengthen its magnitude.

\par To incorporate dynamical electronic correlations beyond the static mean-field treatment of DFT+$U$, we performed DFT+DMFT calculations. The inclusion of frequency-dependent self-energy affects the low-energy electronic structure. Compared to conventional DFT, the quasiparticle bands become narrower and exhibit enhanced spectral weight near the Fermi level. This renormalization increases the sensitivity of the electronic states to the underlying exchange field, leading to a larger effective altermagnetic spin splitting. The DMFT spectral function shown in Fig.~\ref{fig2}(g)-(h) reproduce the key features observed in the DFT+$U$ calculations, namely the enhancement of the momentum-dependent NRASS and the persistence of metallic behavior. The magnitude of NRASS is 0.81 eV  from DFT+DMFT. The agreement between the static DFT+$U$ and dynamical DMFT descriptions indicates that the enhancement of altermagnetic spin splitting is a robust consequence of electronic correlations rather than an artifact of a particular theoretical treatment. The corresponding evolution of magnetic moment from both DFT+U and DFT+DMFT are also presented in table \ref{table-1}.

\par A comparison of the calculated spin splittings obtained from DFT, DFT+$U$, and DFT+DMFT demonstrates a clear correlation-driven trend. The spin splitting increases progressively with increasing local Coulomb interactions. NRASS remains 12.5\% larger in DMFT than the DFT value. It clearly indicates that the enhancement persists even when dynamical electronic correlations are explicitly included. These results establish electronic correlations as a key tuning parameter for controlling altermagnetic properties in CrSb. The correlation-enhanced exchange field substantially amplifies the spin-dependent band separation while preserving the compensated magnetic structure.  Thereby it provides a route toward realizing stronger altermagnetic functionalities in correlated metallic systems. The pronounced sensitivity of the altermagnetic response to electronic correlations suggests that CrSb lies in an intermediate-correlation regime where modest modifications of the effective interaction strength can significantly alter the observable spin splitting. Such tunability may be exploited through chemical substitution, strain engineering, or external pressure, offering promising opportunities for the manipulation of altermagnetic phenomena.

%###################################

\section{Spin-Lattice Coupling as a Route to Enhanced Altermagnetism}

\begin{figure}  
\includegraphics[width=0.5\textwidth,angle=0]{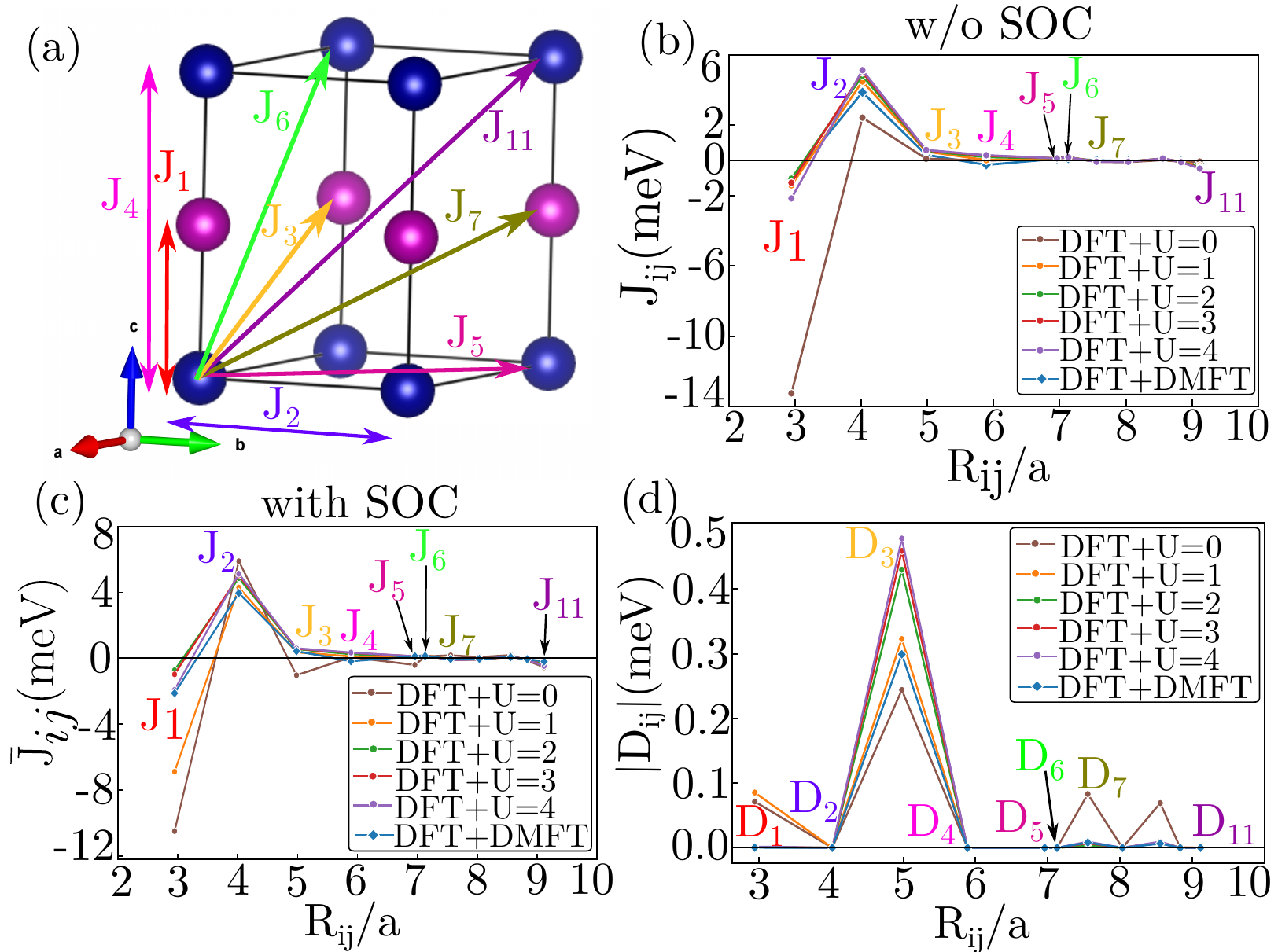} 
\caption{(a) Magnetic exchange interactions for nearest-neighbor (NN) mapping in CrSb. The atoms in blue and green colors represent Cr$_1$ and Cr$_2$ atoms, respectively. Calculated (b)-(c) isotropic exchange part of Heisenberg interactions J$_{ij}$'s without and with spin-orbit coupling (SOC) respectively, (d) antisymmetric Dzyaloshinkii-Moriya interactions D$_{ij}$'s with SOC in CrSb for different NNs. Here R$_{ij}$ (\AA)\ is the NN distance and a is the lattice constant for CrSb..}
\label{fig3} 
\end{figure}

\begin{figure*}  
\includegraphics[width=1.0\textwidth,angle=0]{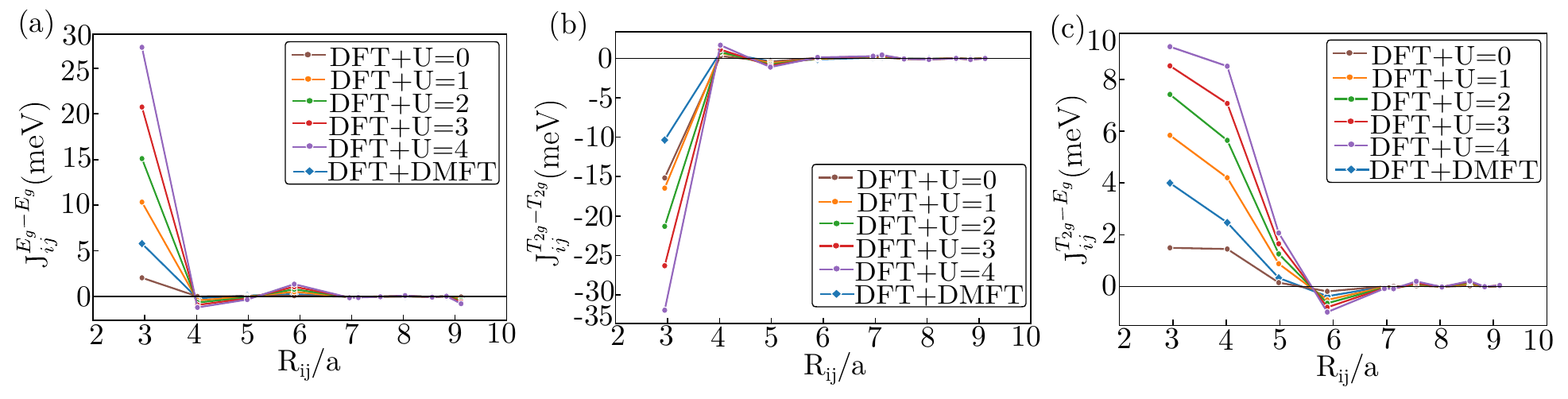} 
\caption{Orbital decomposition of magnetic exchange interactions of Cr-3d  for (a) $J_{ij}^{E_{g}-E_{g}}$, (b) $J_{ij}^{T_{2g}-T_{2g}}$ and (c) $J_{ij}^{T_{2g}-E_{g}}$ as a function of nearest neighbor distance R$_{ij}$/a (\AA) of in CrSb. Here a is the lattice constant of CrSb.}
\label{fig4} 
\end{figure*}

\par SLC quantifies the sensitivity of the altermagnetic state to atomic displacements and phonon-induced lattice vibrations, thereby providing a direct measure of the interplay between magnetic order and lattice dynamics. The electronic structure of AMs is highly sensitive to the underlying magnetic configuration. Therefore understanding the exchange mechanisms responsible for stabilizing noncollinear magnetic order is essential for revealing the role of SLC. To elucidate the microscopic origin of the enhanced NRASS in CrSb, we first investigate the magnetic exchange interactions and their orbital-resolved contributions.

\par Figure~\ref{fig3}(a) depicts the magnetic exchange-interaction network considered for CrSb. The exchange parameters $J_1$, $J_2$, and $J_3$ denote the first, second, and third nearest-neighbor (NN) magnetic interactions, respectively. In total, the crystallographic unit cell includes exchange interactions extending from the first to the seventh NN, as well as the eleventh NN. The remaining exchange interactions arise from magnetic couplings between Cr atoms located in neighboring unit cells. This interaction network provides the basis for evaluating the magnetic exchange parameters and their contributions to the magnetic ground state.

\par The calculated isotropic Heisenberg exchange parameters exhibit a complex interplay between FM and AFM interactions extending beyond the NNs as shown in Figs.~\ref{fig3}(b) and \ref{fig3}(c) for without and with SOC, respectively. The first NN exchange interaction is AFM, while the second NN interaction is FM in nature. Although the overall exchange pattern remains largely unchanged upon the inclusion of SOC, appreciable variations in the magnitudes of the first and second NN exchange interactions indicate the presence of finite SLCs. These dominant exchange interactions govern the stabilization of the NRASS. Their sensitivity to the local structural environment further implies that lattice distortions can effectively modify the magnetic energy landscape.  It consequently influence  the noncollinear magnetic state in presence of SOC. Furthermore, increasing the Hubbard U enhances the magnitudes of both first NN  and second NN exchange interactions (J$_1$ and J$_2$) mainly. It highlights the strengthening of the dominant exchange interactions with increasing electron correlation as illustrated in Figs.~\ref{fig3}(b) and \ref{fig3}(c).

\par The influence of SOC is further evidenced by the emergence of finite DMI, as shown in Fig.~\ref{fig3}(d). Among the various exchange paths, only the first and third NN DMI components are nonzero, and their magnitudes increase gradually with increasing Hubbard U. Here DMIs contribution is coming from
Cr atoms of two different planes (D$_1$ and D$_3$). These antisymmetric exchange interactions favor spin canting and stabilize noncollinear magnetic textures by lowering the energy of spin-spiral configurations, as discussed in the following section. The appearance of finite DMI suggests that structural distortions modify the symmetry of the local exchange pathways, providing an additional mechanism through which lattice degrees of freedom influence the magnetic state. Consequently, SLC not only modifies the isotropic Heisenberg exchange interactions but also affects the noncollinear magnetic ordering.

\par To identify the dominant microscopic exchange channels, we decompose the magnetic interactions into orbital-resolved contributions. As shown in Fig.~\ref{fig4}, the exchange couplings arise from distinct interactions among the Cr-$3d$ orbitals, as obtained from DFT, DFT+$U$, and DMFT calculations. The total exchange interaction in CrSb from the multi-orbital approach can be presented as a sum of three contributions: ${J_{ij}} = {J_{ij}}^{E_{g}-E_{g}} + {J_{ij}}^{T_{2g}-T_{2g}} + {J_{ij}}^{T_{2g}-E_{g}}$.  The $E_g$--$E_g$ channel [Fig.~\ref{fig4}(a)] contributes predominantly to the FM exchange interactions, whereas the $T_{2g}$--$T_{2g}$ channel [Fig.~\ref{fig4}(b)] exhibits a AFM contributions which are increasing with increasing strength of Hubbard U. The mixed $T_{2g}$--$E_g$ interaction [Fig.~\ref{fig4}(c)] is FM in nature and provides an additional source of magnetic frustration. It is crucial for stabilizing noncollinear magnetic states with increasing strength of Hubbard U. Since these orbital-dependent exchange pathways are mediated through Cr--Sb hybridization, even modest lattice distortions can substantially modify their relative strengths.

\par We next investigate the effects of SLC in CrSb using DFT+$U$ calculations for increasing on-site Coulomb interactions, $U=1$-$4$ eV, and compare the results with DMFT. In the presence of lattice vibrations, the magnetic exchange interactions explicitly depends on the atomic displacements. Accordingly, the general form bilinear spin Hamiltonian ${\mathrm{H}_s}$, considering spin-orbit interactions taken into account,  contains symmetric, isotropic Heisenberg exchange, anti-symmetric DMI and symmetric, anisotropic interactions. Therefore, it can be expressed as
\bea
{\mathrm{H}_s} = -\sum_{ijk} \sum_{\{\alpha,\beta\}} e^{\alpha}_i J^{\alpha \beta}_{ij} (\{u_k^{\gamma}\}) e^{\beta}_j 
\label{eq:HamS} \nonumber
\eea
where $e^{\alpha}_i$ ($e^{\beta}_j$) is the $\alpha$ ($\beta$) component of the unitary vector pointing along the direction of the spin located at the site $i$ ($j$). Here  the exchange tensor $J_{ij}^{\alpha\beta}$  become a function of atomic displacements $\{u_k^{\gamma}\}$ due to lattice vibrations as well as the magnetic configuration. The contributions to the mixed spin-lattice Hamiltonian ${\mathrm{H}_{sl}}$ 
can then be obtained by expanding the bilinear magnetic Hamiltonian ${\mathrm{H}_s}$ in displacement. Therefore, the coupled spin-lattice Hamiltonian is given by 
\bea
\label{eq:HamMML}
{\mathrm{H}_{sl}}  = -\sum_{ijk} \sum_{\{\alpha,\beta\}}\Gamma_{ijk}^{\alpha\beta\mu} e_i^{\alpha} e_j^{\beta} u_k^{\mu},  \nonumber
\eea
 where the SLC tensor is defined as
$\Gamma_{ijk}^{\alpha\beta\mu}=\frac{\partial J_{ij}^{\alpha\beta}}{\partial u_k^{\mu}}$. The theory of coupled spin-lattice dynamics has been formulated and successfully applied to various magnetic systems before \cite{PhysRevB.105.104418, SciPostPhys.18.2.064, rb2d-qzqs, PhysRevMaterials.9.024409}. Here, we focus on the effect of SLC on NRASS, which has not been investigated previously to the best of our knowledge. Since altermagnetic spin splitting is fundamentally a non-relativistic phenomenon, we calculate the SLC in the absence of SOC, thereby isolating its intrinsic contribution to NRASS. In this limit, the spin-lattice Hamiltonian (${\mathrm{H}_{sl}}$) contains only the isotropic Heisenberg exchange interaction, and the SLC tensor reduces to the displacement derivative of the exchange interactions given by $\Gamma_{ijk}=\frac{\partial J_{ij}}{\partial u_k} $.

\begin{figure}  
\includegraphics[width=0.5\textwidth,angle=0]{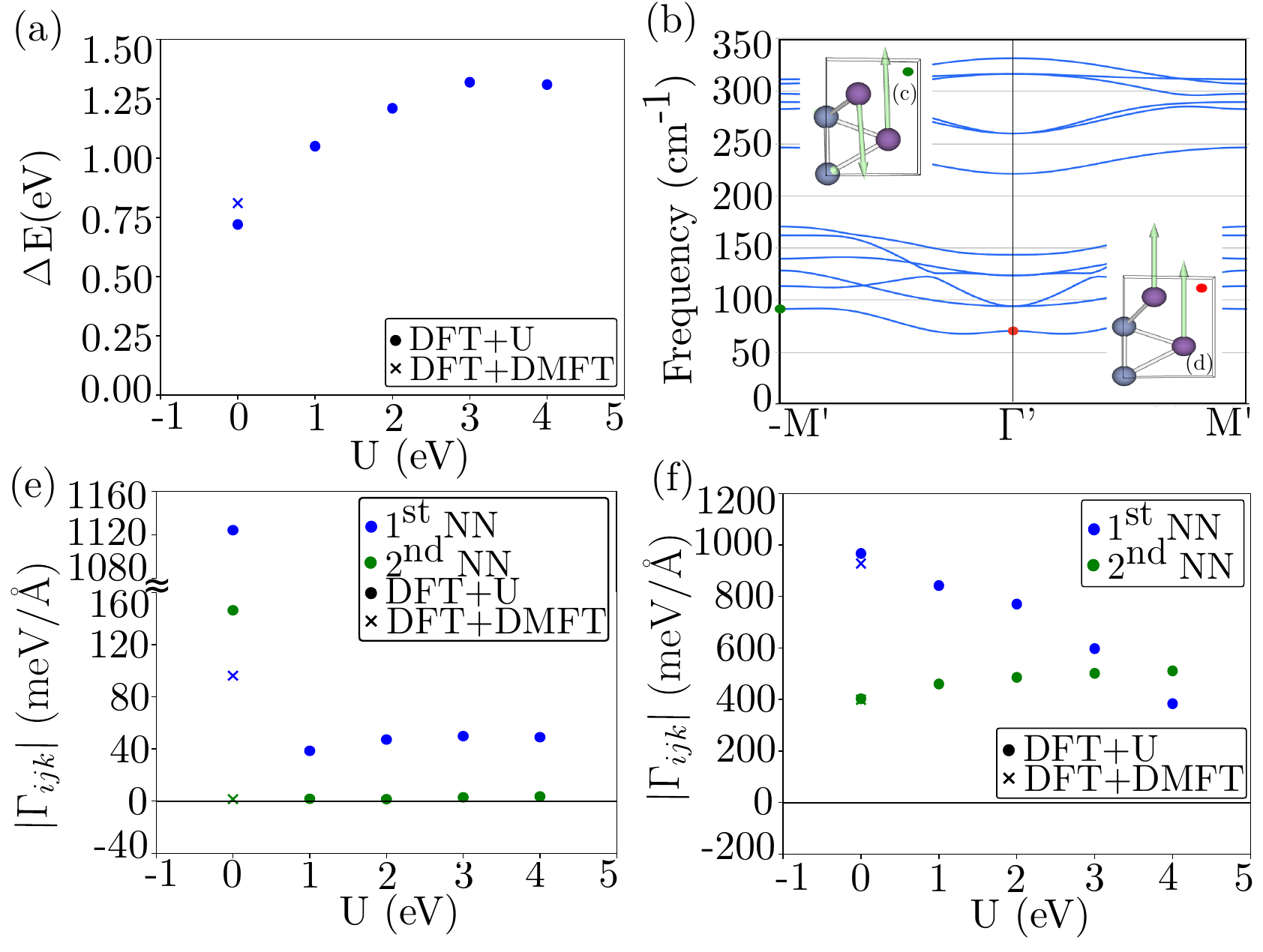} 
\caption{(a) Variation of the non-relativistic altermagnetic spin splitting (NRASS) with the Hubbard parameter U, and (b) phonon band structure for CrSb. The thermal lattice vibrations of Sb atoms occur in the (c) opposite and (d) same directions at the $M'$ and $\Gamma'$ points, respectively where the gray and violet spheres represent Cr and Sb atoms, respectively. The calculated spin-lattice couplings as a function of U for (e) in-plane displacements of Cr atoms along the Cr–Cr bonds and (f) out of plane displacements of Sb atoms at the $\Gamma'$ point.}
\label{fig5} 
\end{figure}

\par The correlation-induced enhancement of NRASS in CrSb is shown in Fig.~\ref{fig5}(a). This enhancement indicates an important role of the coupling between the spin and lattice degrees of freedom. To examine the lattice contribution, we analyze the phonon modes involving both Cr and Sb atoms. We consider the $M'$-$\Gamma'$-$M'$ path in the BZ, where a sizable altermagnetic spin splitting is observed. The atomic displacements associated with the phonon eigen modes are analyzed using the interactive phonon visualizer toolkit \cite{phonon}. Figures~\ref{fig5}(b)-(d) show the phonon dispersion and representative phonon eigenvectors at the high-symmetry $\Gamma'$ and $M'$ points where arrows show the vectors of oscillation due to lattice vibration individual atoms. At the $\Gamma'$ and $M'$ points, the Cr atoms show both in-plane and out of plane oscillations. Both are mainly along the Cr-Cr bond directions. Whereas, the Sb atoms exhibit predominantly out of plane oscillations. Depending on the phonon mode, neighboring Sb atoms oscillates either in the same or opposite directions, as shown in Figs.~\ref{fig5}(c) and \ref{fig5}(d). The phonon modes away from the $\Gamma'$ and $M'$ points involve more complex oscillations of both Cr and Sb atoms. We take the oscillation directions of the Cr and Sb atoms obtained from the phonon eigenvectors as the corresponding displacement directions for calculating the SLC. We observed that the displacement amplitudes are generally larger for Sb than for Cr. This indicates stronger lattice modulation of the Sb sublattice than the Cr sublattice highlighting the significant role of the Sb sublattice environment in modulating the altermagnetic response of CrSb. Therefore, the Sb atoms are expected to make a larger contribution to the SLC than the Cr atoms. This enhanced contribution may play an important role in the correlation induced modification of NRASS.

\par Figure~\ref{fig5}(e) and (f) show the calculated SLCs for in-plane and out of plane displacements of the Cr and Sb atoms, respectively, at the $\Gamma'$ point. Here we use displacement magnitude $\Delta$U = 0.01 \AA\ for both Cr and Sb atoms for calculating SLC which is in the linear region of calculated SLC with displacements. The SLCs associated with Sb displacements are approximately one order of magnitude larger than those associated with Cr displacements. Thus, the Sb sublattice provides the dominant contribution to the SLC in CrSb. For Cr displacements, the calculated SLCs for the first and second NN exchange interactions are -1124.47~meV\AA\ and -146.32 ~meV/\AA, respectively. These values are substantially larger than the corresponding DMFT results of 96.18~meV/\AA\ and -1.53~meV/\AA. The SLCs for the first and second NN interactions increase systematically with increasing Hubbard interaction $U$. This trend closely follows the correlation-induced enhancement of NRASS for Cr displacements in CrSb. The first NN exchange interaction gives the dominant contribution to the SLC. Its strength also increases with increasing $U$. The calculated SLCs for displacements of the Cr and Sb atoms at the $M'$ point are displayed in Appendix.

\par In contrast, the SLC associated with Sb displacements decreases with increasing $U$. This opposite $U$ dependence of the Cr- and Sb-mediated SLCs highlights their distinct roles in the correlation dependence of NRASS. The structural distortions modify the Cr-Sb bond geometry and the local crystal-field environment. They consequently alter the orbital hybridization and the underlying exchange interactions. These changes modify the balance between competing exchange channels and enhance the exchange frustration. The resulting modification of the exchange interactions amplifies the momentum-dependent spin splitting characteristic of altermagnetism. These results establish SLC as a microscopic mechanism connecting electronic correlations to the enhanced altermagnetic response in CrSb.

\begin{table}
\centering
\begin{tabular}{|c|c|c|c|c|}
\hline
Calculation & Magnetic moment($\mu$) & $\Delta$E$_1$ & $\Delta$E$_2$ \\
\hline
DFT & 3.47 & 0.19 & 1.18 \\
\hline
DFT+U=1 & 3.87 & 0.22 & 1.17 \\
\hline
DFT+U=2 & 4.10 & 0.25 & 1.06 \\
\hline
DFT+U=3 & 4.24 & 0.30 & 1.01 \\
\hline
DFT+U=4 & 4.33 & 0.37 & 0.92 \\
\hline
DFT+DMFT & 3.47 & 0.22 & 1.15 \\
\hline
\end{tabular}
\caption{Calculated magnetic moment and maximum values of the two non-relativistic altermagnetic spin splittings (NRASSs) along the -D–P–D direction in the Brillouin zone for MnTe, obtained from DFT+$U$ with $U=1$–4 eV and from DMFT calculations.}
\label{table-2}
\end{table}

\begin{figure}  
\includegraphics[width=0.5\textwidth,angle=0]{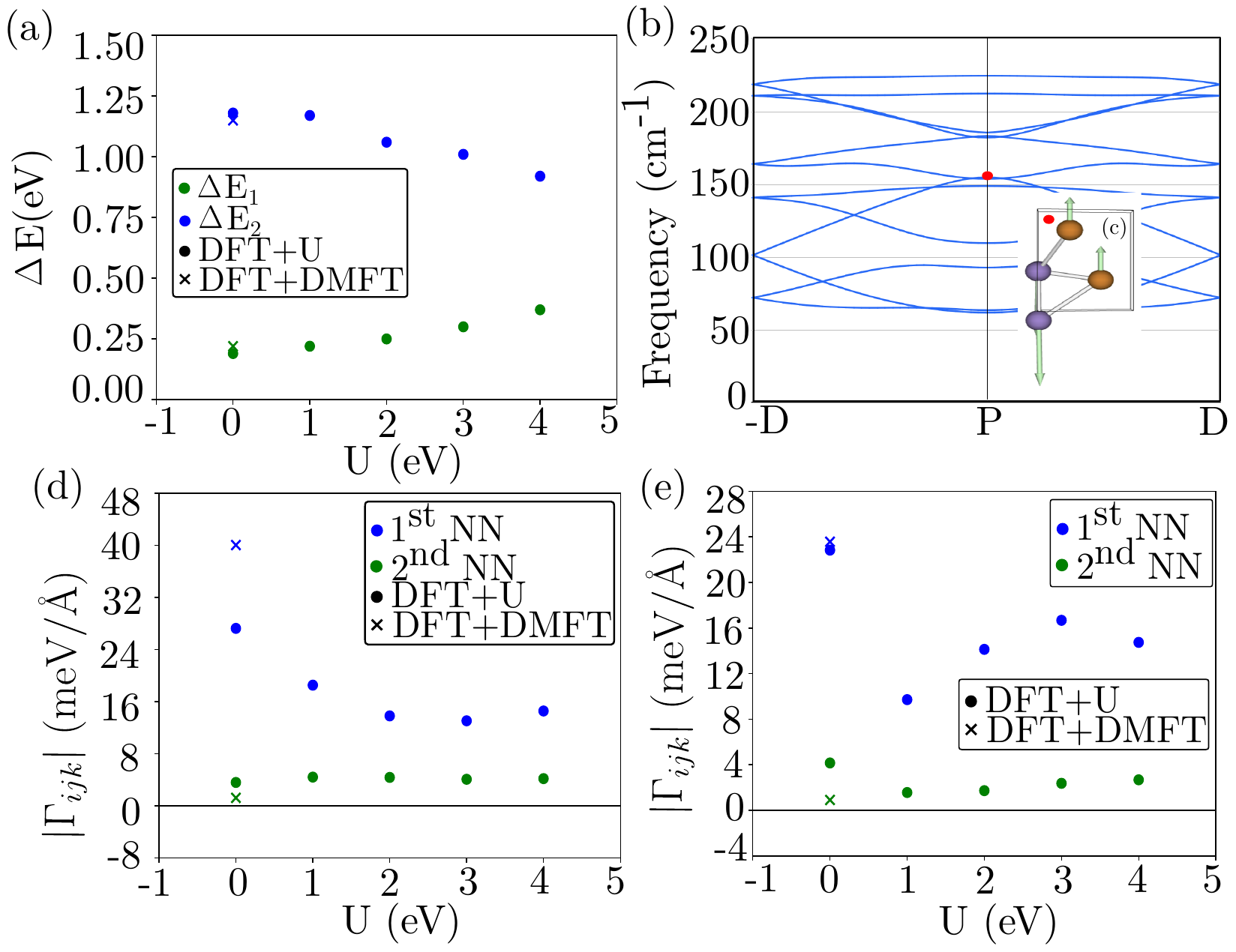} 
\caption{(a) Variation of the non-relativistic altermagnetic spin splitting (NRASS) with the Hubbard parameter U, and (b) phonon band structure for MnTe. (c) The thermal lattice vibration for Mn and Te atoms at P point where the violet and orange spheres represent Mn and Te atoms, respectively. The calculated spin-lattice couplings as a function of U for out of plane displacements of (d) Mn atoms along the Mn–Mn bonds, and (e) Te atoms along z direction at the P point respectively. }
\label{fig6} 
\end{figure}

\par To further examine the generality of this mechanism, we investigate SLC in another NiAs-type altermagnet, MnTe, which exhibits two distinct NRASS channels (see Appendix). These two channels show opposite dependencies on the Hubbard interaction $U$, as shown in Fig.~\ref{fig5}(d) (see also Table~\ref{table-2}). $\Delta E_1$ increases with increasing $U$, whereas $\Delta E_2$ decreases. Since $\Delta E_2$ provides the dominant contribution to the total NRASS, the net altermagnetic spin splitting is progressively suppressed with increasing $U$. We evaluate the SLC associated with the first and second NN exchange interactions, which provide the dominant contributions to the magnetic exchange in MnTe. The corresponding orbital-resolved decomposition of these exchange interactions is presented in the Appendix.

\par Figures~\ref{fig6}(b)-(c) show the calculated phonon modes and the corresponding atomic displacement patterns at the high-symmetry $P$ point. The Mn atoms exhibit a larger out of plane displacement, along the Mn-Mn bond direction, than the Te atoms. This difference in the oscillation amplitudes of corresponding phonon modes is directly reflected in the magnitude of the SLC.  The calculated SLC induced by Mn displacement is substantially stronger than that induced by Te displacement as shown in Figs.~\ref{fig6}(d) and 6(e). Interestingly, the $U$ dependence of the SLC also exhibits a clear sublattice-dependent behavior. The SLC associated with Mn displacement decreases with increasing $U$, following the same trend as the dominant NRASS channel $\Delta E_2$. In contrast, the SLC associated with Te displacement increases with $U$, exhibiting the opposite trend. This contrasting behavior demonstrates that the Hubbard interaction modifies the altermagnetic spin splitting through distinct lattice-coupling channels associated with different atomic sublattices. This demonstrates that SLC serves as a useful microscopic descriptor for understanding and tuning NRASS in altermagnets. More broadly, the interplay among lattice distortions, exchange interactions, and altermagnetic spin splitting suggests that lattice-engineering strategies, including strain, pressure, phonon excitation, and chemical substitution, could provide effective routes for controlling altermagnetic responses.

%###################################

\begin{figure}  
\includegraphics[width=0.5\textwidth,angle=0]{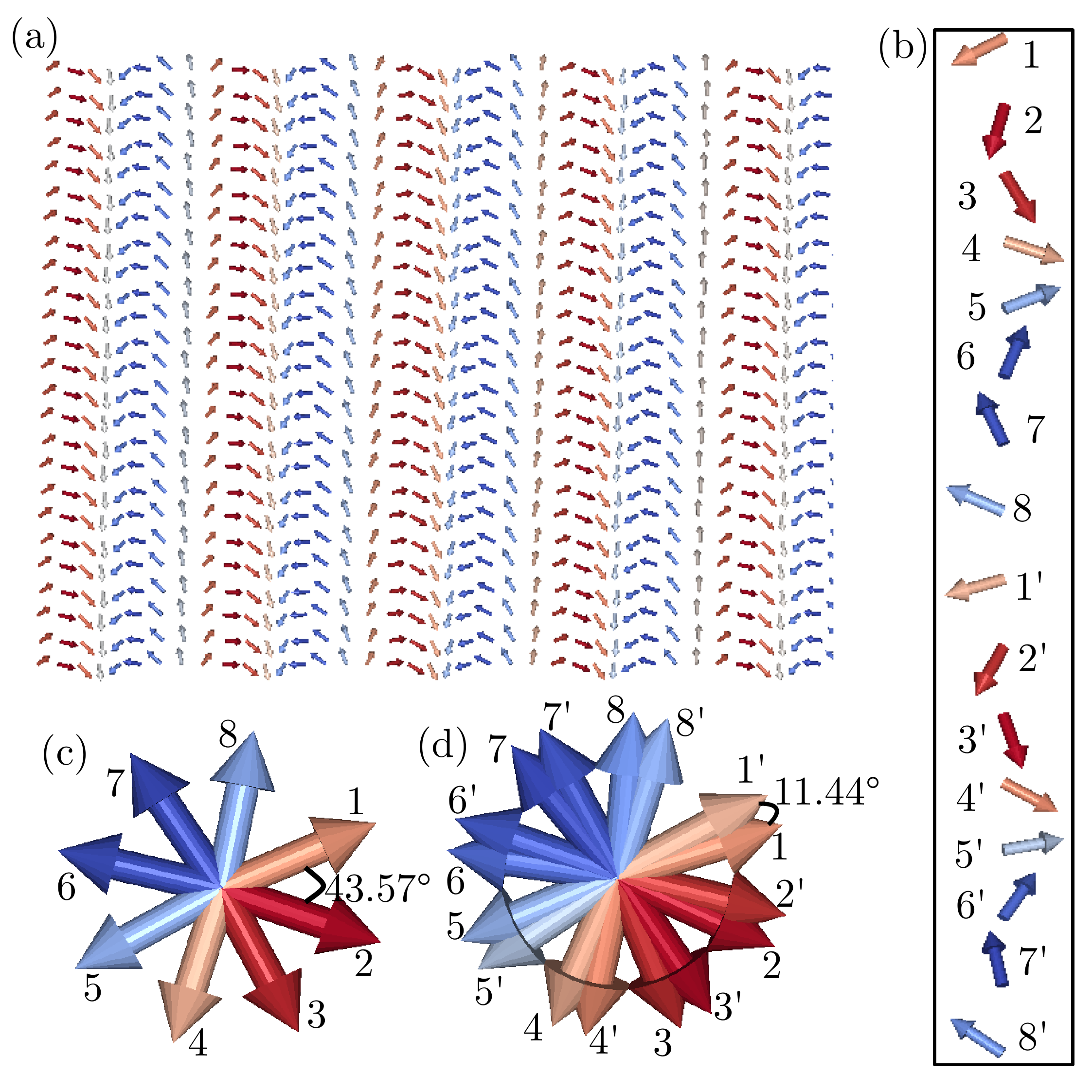} 
\caption{Spin-spiral structure of CrSb obtained from DFT+$U$ with $U=2$ eV. (a) Projection onto the $YZ$ plane, (b) side view, and (c)–(d) top views illustrating the pitch angle of $43.57^\circ$ and an incommensurate spiral angle of $11.44^\circ$.}
\label{fig7} 
\end{figure}

\section{Correlation-mediated evolution of spin-spiral order}

\par We further investigate the evolution of the magnetic ground state as a function of electronic correlations in the presence of SOC. Our calculations reveal a correlation-driven transition from a collinear AFM state to an incommensurate spin-spiral (SS) state, demonstrating that electronic correlations play a decisive role in stabilizing noncollinear magnetism in CrSb. Within conventional DFT ($U=0$) and DFT+$U$ for $U<2$ eV, the collinear AFM configuration remains energetically favorable. In this regime, neighboring Cr moments are aligned antiparallel, resulting in a fully compensated magnetic structure with zero net magnetization. Upon increasing the on-site Coulomb interaction, the balance among competing exchange interactions is progressively modified. For $U\geq2$ eV, the collinear AFM state becomes unstable, and an incommensurate SS configuration is energetically favored. This correlation-driven magnetic instability originates from the enhanced competition between the first NN AFM exchange and second NN FM exchange interactions in CrSb, whose relative strengths are substantially modified by correlation-induced changes in the Cr-$3d$ orbital hybridization and corresponding orbital resolved magnetic exchanges.

\par Figure~\ref{fig7} shows the spin texture of the incommensurate SS state obtained for $U=2$ eV. The spiral propagates along the $z$ direction and exhibits a periodicity of eight Cr layers, corresponding to a pitch angle of $43.57^\circ$ and an incommensurate angle of $11.44^\circ$, as shown in Figs.~\ref{fig7}(b)-(d). With increasing $U$, the spiral wavelength progressively decreases, accompanied by an increase in the pitch and incommensurate angles, indicating the stabilization of a more strongly twisted magnetic texture. To assess the robustness of this correlation-driven noncollinear state beyond the static mean-field description, we further performed DFT+DMFT calculations. The DMFT calculations likewise stabilize an incommensurate SS state, with a periodicity of seven Cr layers, a pitch angle of $50.09^\circ$, and an incommensurate angle of $9.07^\circ$ (see Appendix).  These results establish electronic correlations favor noncollinear magnetic order with enhanced altermagnetic response in CrSb.

%#################################

\section{Conclusion}

\par In summary, we have investigated the interplay between electronic correlations and SLC in the NiAs-type altermagnet CrSb using DFT, DFT+$U$, and DMFT. We find that increasing electronic correlations systematically enhances NRASS, while dynamical correlations further modify the spin splitting through the renormalization of the Cr-$3d$ electronic states. By explicitly evaluating the SLC, we establish a direct correspondence between its evolution and that of the dominant NRASS which demonstrates that the lattice sensitivity of the magnetic exchange interactions plays a central role in the correlation-dependent altermagnetic response. A comparative study of the NiAs-type altermagnet MnTe reveals a similar relationship between the dominant exchange interactions, their SLC, and the corresponding NRASS, suggesting that this mechanism extends beyond CrSb. Furthermore, electronic correlations drive the stabilization of noncollinear SS phase in CrSb while simultaneously enhancing its altermagnetic spin splitting. Overall, our results establish SLC as an important microscopic descriptor of the strength and tunability of altermagnetic spin splitting. Itb provide a framework for controlling altermagnetic responses through lattice-engineering strategies, including strain, pressure, phonon excitation, and chemical substitution.

\begin{acknowledgments}
BS  thanks the Prime Minister’s Early
Career Research Grant (PMECRG) of the Anusandhan National Research Foundation Grant No. ANRF/ECRG/2024/005021/PMS. 
\end{acknowledgments}

\bibliography{crsb-sd}

%#################

\appendix
\section*{Appendix}
\label{appendix}

\par
\begin{figure*}  
\includegraphics[width=0.6\textwidth,angle=0]{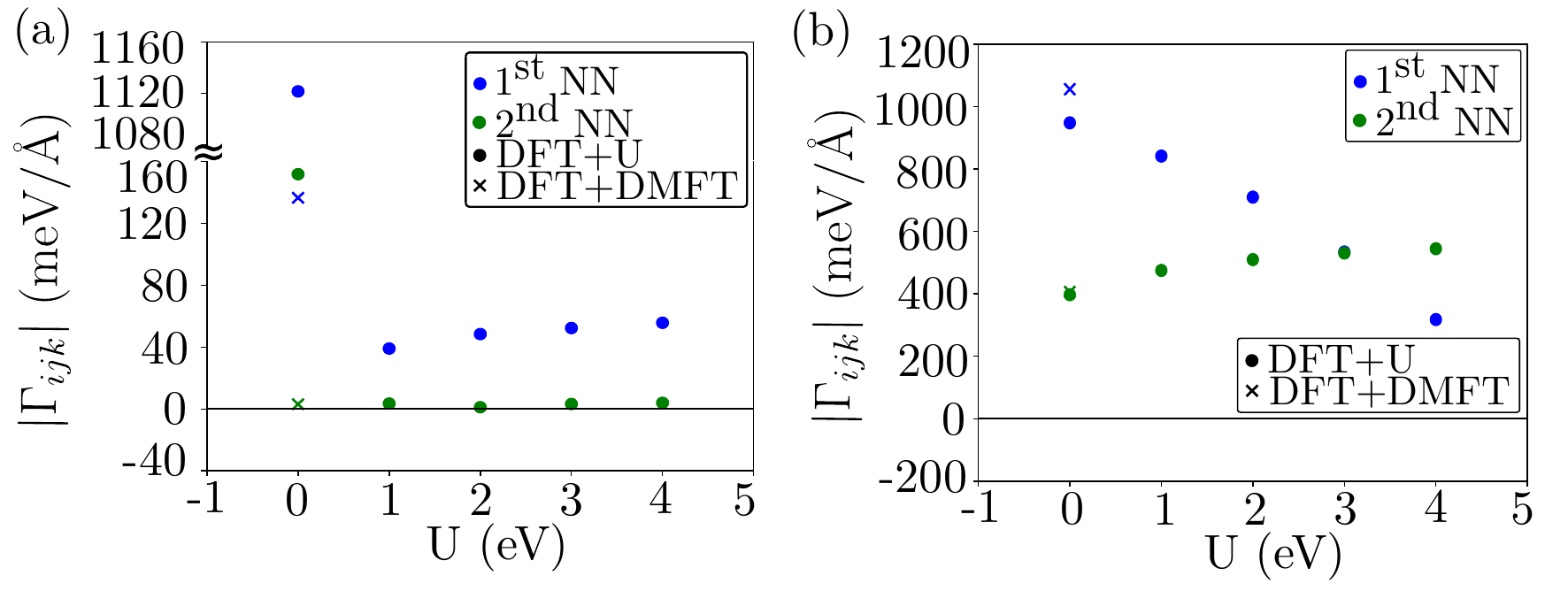} 
\caption{The calculated spin-lattice couplings as a function of U for out of plane displacements of (a) Cr atoms along the Cr–Cr bonds and (b) Sb atoms along z direction at the high symmetry $M'$ point where the thermal lattice vibrations of Sb atoms occur in opposite directions.}
\label{sfig1} 
\end{figure*}

\begin{figure*}  
\includegraphics[width=0.96\textwidth,angle=0]{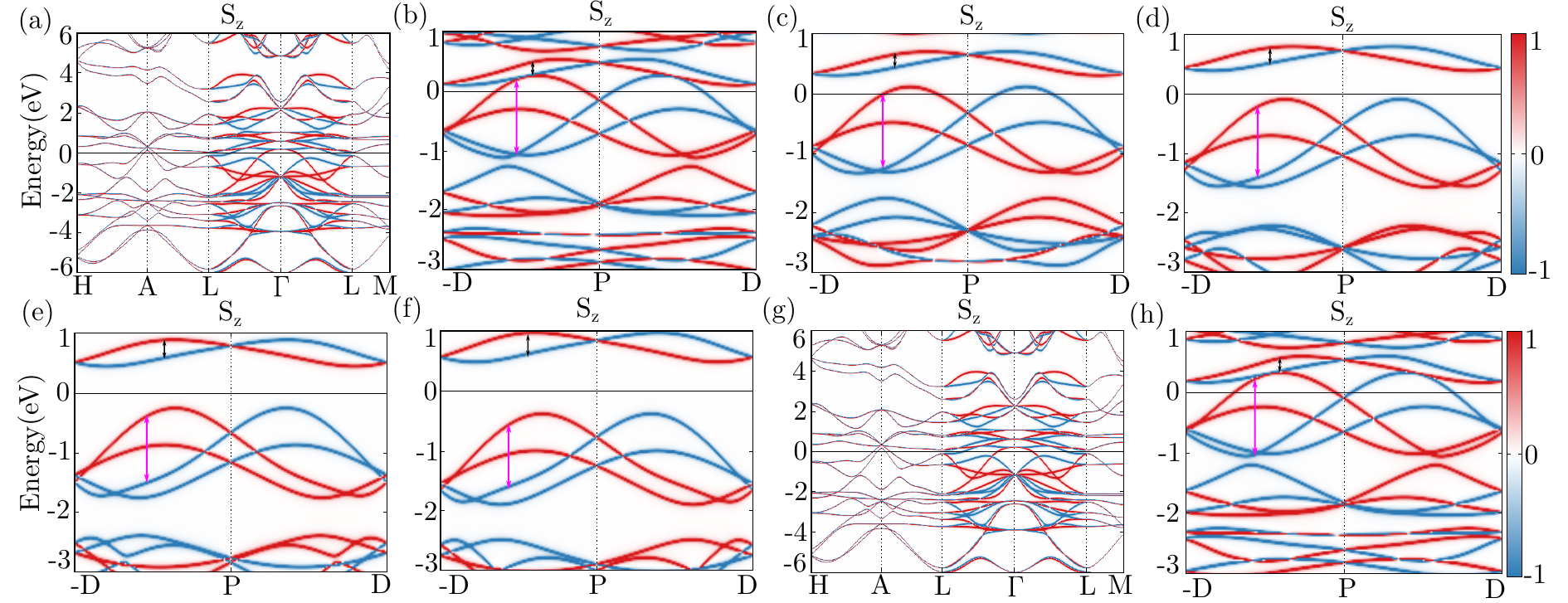} 
\caption{Non-relativistic spin-split band structures for MnTe without SOC along the H-A-L-$\Gamma$-L-M and -D-P-D paths from DFT+$U$ with (a)–(b) $U=0$, (c) $U=1$, (d) $U=2$, (e) $U=3$, and (f) $U=4$ eV, and from (g)–(h) DFT+DMFT respectively. Here the black and pink arrows show two spin-splitting channels $\Delta$E$_1$ and $\Delta$E$_2$ respectively.}
\label{sfig2} 
\end{figure*}

\begin{figure*}  
\includegraphics[width=1.05\textwidth,angle=0]{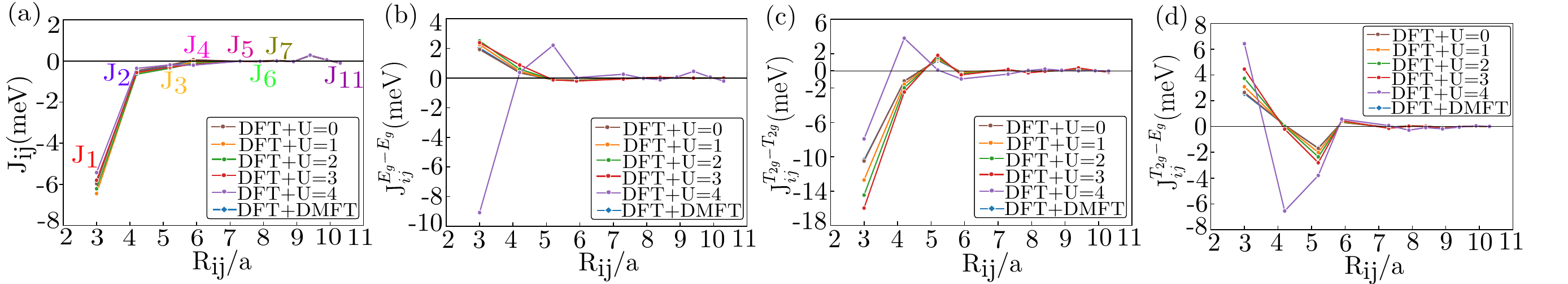} 
\caption{Calculated (a) $J_{ij}$ without SOC and  its orbital decomposition for Mn-3d (b) $J_{ij}^{E_{g}-E_{g}}$, (c) $J_{ij}^{T_{2g}-T_{2g}}$ and (d) $J_{ij}^{T_{2g}-E_{g}}$ contributions as a function of nearest neighbor distance R$_{ij}$/a (\AA) in MnTe from DFT+U and DFT+DMFT calculations. Here a is the lattice constant for MnTe.}
\label{sfig3} 
\end{figure*}

\begin{figure}  
\includegraphics[width=0.5\textwidth,angle=0]{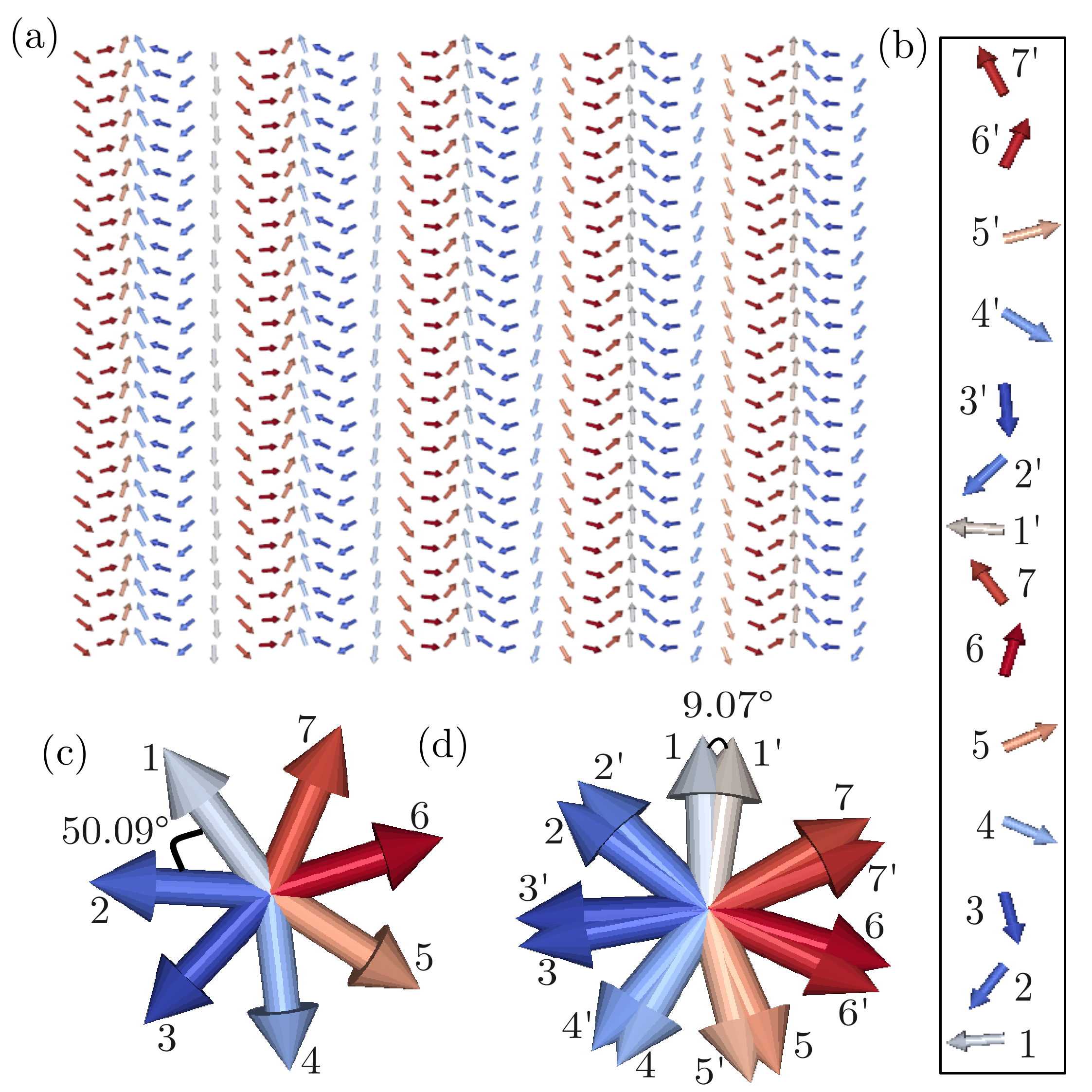} 
\caption{Spin-spiral structure of CrSb obtained from DFT+DMFT. (a) Projection onto the $YZ$ plane, (b) side view, and (c)–(d) top views illustrating the pitch angle of $50.09^\circ$ and an incommensurate spiral angle of $9.07^\circ$.}
\label{sfig4} 
\end{figure}

 Figure \ref{sfig1}(a)-(b) show the calculated spin-lattice coupling constants as a function of U for out of plane displacements of both of Cr and Sb atoms respectively at high symmetry $M'$ point. Figure \ref{sfig2}(a)-(h) present the calculated band structures along the paths H-A-L-$\Gamma$-L-M and -D-P-D for MnTe from DFT+U with  $U=1$--$4$ eV and from DFT+DMFT calculations respectively. Figure \ref{sfig3}(a)-(d) show the J${_{ij}}$ without spin-orbit coupling (SOC) and its orbital contributions for Mn-3d orbital to E$_g$-E$_g$, T$_{2g}$-T$_{2g}$ and T$_{2g}$-E$_g$ channels in MnTe as a function of the neatest neighbor distance R$_{ij}$ (\AA) from both DFT+U and DFT+DMFT. Figure \ref{sfig4}(a)-(d) display the spin texture of CrSb obtained by combining spin-dynamics and Monte Carlo simulations and using the magnetic exchange interactions calculated within DFT+DMFT including SOC.

\end{document}